\documentclass[prl,twocolumn,showpacs,floatfix,amsfonts]{revtex4}
\usepackage{graphicx,graphics,color,epsfig}
\usepackage{bm}
\usepackage{amsmath}
\usepackage{amssymb}

\newcommand{\bk}{{\bf k}}

\newcommand{\br}{{\bf r}}

\newcommand{\beqa}{\begin{eqnarray}}
\newcommand{\eeqa}{\end{eqnarray}}

\begin{document}
\title{Detecting Breather Excitations with Inelastic Tunneling
Spectroscopy}
\author{Jian-Xin Zhu}
\author{K. \O. Rasmussen}
\author{A. R. Bishop}
\author{A. V. Balatsky}
\affiliation{Theoretical Division, Los Alamos National Laboratory,
Los Alamos, New Mexico 87545}
\date{\today}
\begin{abstract}
We propose inelastic electron tunneling 
spectroscopy scanning tunneling microscopy (IETS-STM) as a means of 
exciting and observing intrinsic
localized modes (breathers) in a  macromolecule. As a demonstration, inelastic
tunneling features of the density of states
are calculated for a simple nonlinear elastic Morse chain. The general 
formalism we have developed for the IETS is
applicable to other nonlinear extended objects, such as DNA
on a substrate.
\end{abstract}
\maketitle
 
Dynamical localization phenomena have been a subject of intense
theoretical research since discrete breathers (or intrinsic
localized modes) were proven \cite{Mackay_Aubry} to be generic
solutions of nonlinear spatially discrete systems. These solutions of
the underlying equations of coupled nonlinear oscillators are
characterized by being temporally periodic and spatially 
localized~\cite{Flach,Aubry,Cam}. Two particularly important and
physically appealing aspects of discrete breathers are 1) that
they exist in translationally invariant systems and therefore
these excitations are radically different from Anderson
localization and other localization effects driven by external disorder or defects, and
2) that they are very robust with respect to changes in the
equation of motion. They have been discussed in such diverse systems
as spin lattices \cite{Sievers}, and dynamics of DNA~\cite{Peyrard,NAR}.
 
The robust localization in the form of discrete breathers occurs
because the discreteness of the system provides an effective
cutoff for the wavelength of the linear modes (e.g. phonons) 
that may exist in the
system, and this can prohibit phonon resonances with all
temporal harmonics of the discrete breather. This effectively
eliminates decay mechanisms into extended linear modes.
It is the nonlinearity of the systems that makes this possible by
admitting frequencies outside the phonon (linear) band.
 
The field has been dominated by theoretical research, and
general methods for direct experimental observation of breathers are lacking,
although experiments on localization of light propagation in
weakly coupled optical waveguides \cite{Eisenberg},
low-dimensional crystals \cite{Swanson} and Josephson-junction
networks \cite{Flach_JJ,Orlando} have recently been reported.

Here we propose the use of scanning tunneling
microscopy (STM) to both excite and observe breathers in an inelastic
electron tunneling spectroscopy experiment (IETS). The IETS itself is
a well established and powerful tool that allows the measurement
of the  characteristic energies of local and extended modes.
Examples of applications of this technique, among many, include
measurements of molecular stretching and vibrational modes in
metal-insulator-metal tunnel junctions~\cite{Jakievich1,Hansma1};
local IETS scanning tunneling microscopy (IETS-STM)
measurements of the local vibrational mode of  {\em single} atoms
or molecules on a surface~\cite{Ho1}; observation in
tunneling of the collective magnetic resonance in the
superconducting state of high-$T_c$ 
materials~\cite{Norman1,Zasadzinski1}; and  recent IETS-STM of the spin flip
of a magnetic atom in a magnetic field~\cite{Heinrich1}. Here we 
give for the first time a theoretical basis for an IETS-STM observation
of the breather mode in an extended system.  

\begin{figure}[th]
\centerline{\psfig{file=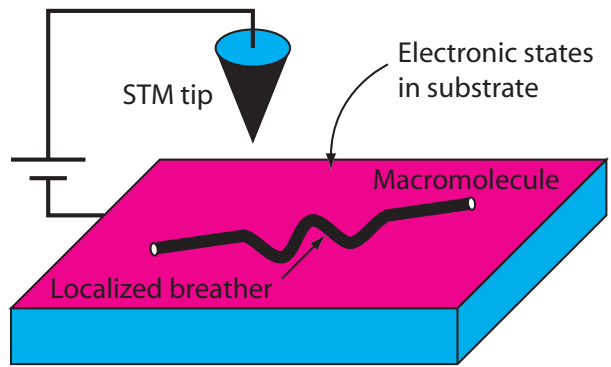,width=8cm}} \caption{(Color) The proposed
experimental realization: An extended molecule is placed
on the metallic surface. An STM is used to inject/extract electrons
from the molecule. Electron-vibration coupling produces 
excitations that are outside of the phonon continuum. This 
localized excitation will re-form into a set
of breather excitations with different energies. The breathers
will remain localized for a long period of time since they
cannot decay into the phonon continuum. Subsequent IETS will
reveal localized excitations.} \label{FIG:SETUP}
\end{figure}

The IETS directly measures excitation energies. Electrons scatter off
a local (or collective) mode. Due to the scattering, a
contribution to the electron self-energy occurs above a
corresponding threshold value of the frequency determined by the
mode frequency. Thus, in a tunneling experiment, for a voltage bias
exceeding the threshold, electrons can excite the mode. This
additional scattering channel leads to a step in the density of
states and in the tunneling conductance. Low temperatures
are required to avoid thermal smearing of the step in the
conductance.

The setup we propose is to use STM in a two step process: First
we propose to locally excite breathers  by injecting energetic
electrons into the molecule on the substrate.  The energy of the
injected electrons is taken to be outside the phonon band and able
to excite the breather; in the subsequent step we propose to
measure the local density of states (LDOS) in the vicinity of the
injection point. The first step is a major assumption of this work. 
There is no detailed microscopic theory for the energy transfer
between electrons and breathers. 
We make the reasonable assumption that
the resonance between the energy of the electron and that of the breather
 will allow efficient energy transfer between electronic
degrees of freedom and breathers. At the second stage, the 
LDOS at a finite bias but with much smaller drive currents will
allow measurement of the inelastic processes revealing the
existence of the localized breather. Alternatively, one can view
our work as addressing what will be the features in LDOS once the
breathers are excited. The setup is illustrated in Fig.~\ref{FIG:SETUP}.

In what follows we will use a simple model to
demonstrate the possibility of IETS in macromolecules. We 
consider a translationally invariant long chain with vibrational
sites $n =-N+1,\dots,0,\dots,N$
 with local molecular displacements $x_n$ at each site, 
for simplicity taken only along the direction normal to the
surface. The role of the surface in this consideration is to
provide mechanical support for the chain and to provide conduction
electrons whose DOS will ultimately be measured by STM.

Specifically, we consider the following model Hamiltonian: 
\beqa
H &=& \sum_{n = -N+1 }^{n =N} \biggl{[}\frac{M\dot{x}_n^2}{2} + V(x_n) +
\frac{k}{2}(x_n - x_{n-1})^2\biggr{]} \nonumber \\
&&+ \sum_{\mathbf{k}}\xi_{\mathbf{k}}
c_{\mathbf{k}}^{\dagger}c_{\mathbf{k}} + H_{int} \;.
\label{EQ:Ham1} 
\eeqa 
Here the first three terms describes the dynamics for the displacement field
in an extended system, the fourth term is the Hamiltonian for uncoupled surface
electrons with $\xi_{\mathbf{k}}=\epsilon_{\mathbf{k}}-\mu$ the energy 
dispersion measured with respect to the chemical potential, the last term 
$H_{int}  =\lambda \sum_{n=-N+1}^{n=N}  x_n c^{\dag}_n c_n $ is
the interaction Hamiltonian that describes a local (Holstein)
coupling between the displacement $x_n$ and local electronic
density at a site $n$. We will ignore spin indices of the fermion
operators $c_n$ ($c^{\dag}_n$), since we are interested in a total
electronic density and there is no explicit spin dependence in the
Hamiltonian.
 
We introduce the Matsubara Green's functions of the electrons on
the substrate. The bare Green's function is given
by:
 \beqa
  G^0(\mathbf{k},i\omega_{n}) &=& \frac{1}{i\omega_{n} -
  \xi(\mathbf{k})}\;,
 \\ G^0(\mathbf{r},i\omega_{n})
 &=& \int \frac{d^dk}{(2\pi)^{d}}
e^{i\bk \cdot \br}G^0(\bk,i\omega_{n})\;, \label{eq:Gelectron1}
\eeqa where $\omega_{n}=(2n+1) \pi T$ is the Matsubara frequency
for electrons. In a two-dimensional (2D) system, assuming that the
surface states of the electrons are interacting strongly with the
displacements, we have $G^0(\br,i\omega_{n}) =-i \pi N_0
\mbox{sgn}(\omega_{n}) J_0(k_F r)$, with $N_0$ being the density of
states of the metal,  $J_0(x)$ the Bessel function of the zeroth
order, and $\mbox{sgn}(x)$ the sign function. Employing
perturbation theory up to second order (in the coupling constant
$\lambda$) in the presence of the dynamical disorder (discrete
breathers) coupled to the electrons, the full Green's function is
\begin{eqnarray}
G(\mathbf{r},\mathbf{r}^{\prime};i\omega_{n}) & = &
G^{0}(\mathbf{r},\mathbf{r}^{\prime};i\omega_{n}) + \int\int d\mathbf{r}_{1}
d\mathbf{r}_{2} G^{0}(\mathbf{r},\mathbf{r}_{1};i\omega_{n})
\nonumber \\
&& \times \Sigma (\mathbf{r}_{1},\mathbf{r}_{2};i\omega_{n}) G^{0}
(\mathbf{r}_{2}, \mathbf{r}^{\prime};i\omega_{n})\;,
\label{EQ:FULL_GREEN}
\end{eqnarray}
with self-energy
\begin{eqnarray}
\Sigma(\mathbf{r}_{1},\mathbf{r}_{2};i\omega_{n})&=&\frac{\lambda^{2}}{
\beta} \sum_{i\Omega_{m}}
D^{0}(\mathbf{r}_{1},\mathbf{r}_{2};i\Omega_{m})
\nonumber \\
&&\times
G^{0}(\mathbf{r}_{1},\mathbf{r}_{2};i(\omega_{n}-\Omega_{m}))\nonumber
\\
&=& \lambda^{2}\int \int \frac{d^{2}k d\omega}{(2\pi)^{2} }
\frac{n_{F}(\xi_{\mathbf{k}})-1-n_{B}(\omega)}{i\omega_{n}
-\xi_{\mathbf{k}}
-\omega} \nonumber \\
&& \times B(\mathbf{r}_{1},\mathbf{r}_{2};\omega)\;.
\label{EQ:SELF_ENERGY}
\end{eqnarray}
Here $D^{0}(\mathbf{r}_{1},\mathbf{r}_{2};i\Omega_{m})$ is the
Fourier transform of the correlation function
$D^{0}(\mathbf{r}_{1},\mathbf{r}_{2}; \tau_{1}-\tau_{2})=\langle
T_{\tau} [x(\mathbf{r}_{1}\tau_{1})
x(\mathbf{r}_{2}\tau_{2})]\rangle$, $\Omega_{m}=2m\pi T$ is the
Matsubara frequency for the bosonic excitations, and the two
distribution functions are given by $n_{F,B}(E)=1/[\exp(E/T)\pm
1]$, respectively. Of great importance in
Eq.~(\ref{EQ:SELF_ENERGY}) is the {\em generalized spectral
function} $B(\mathbf{r}_{1},\mathbf{r}_{2};\omega)$, which has not yet
been specified and depends on the detailed model for the
bosonic excitations. In this sense, Eq.~(\ref{EQ:FULL_GREEN})
together with Eq.~(\ref{EQ:SELF_ENERGY}) is the most general
formalism to describe the effects of the bosonic excitations on
the electronic properties.

\begin{figure}[th]
\centerline{\psfig{file=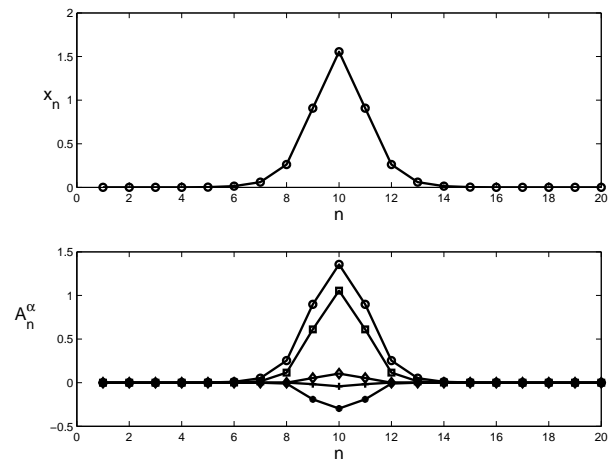,width=8cm}} \caption{The
  displacement field in the breather
excitation (a), and its Fourier transform b) $\omega_0=0.7$, $A_n^0$ - squares, $A_n^1$ -
circles, $A_n^2$ - small circles, $A_n^3$ - diamonds, $A_n^4$ -
pluses} \label{FIG:XDISTR}
\end{figure}

For the purpose of the present work, we need to identify the
spectral function $B(\mathbf{r}_{1},\mathbf{r}_{2};\omega)$ for
the breather mode.  Since the dynamics for the breather in 
(\ref{EQ:Ham1}) is
described through a lattice model, we are actually studying the
correlation function in discretized space coordinates. In real
time and space, the retarded correlation function is given by:
\beqa D^{0,r}(n,m;t,t') = i\theta(t-t^{\prime}) \langle
[\hat{x}_n(t), \hat{x}_m(t')]\rangle\;, \label{eq:Gx1} \eeqa 
where $n, m \in \mbox{\em molecule sites}$, i.e., $\mathbf{r}_{n}$ 
and $\mathbf{r}_{m}$,  and
$\hat{x}_n(t)$ is the displacement operator at the site
 $n$. We use  standard quantized notations: 
 \beqa \hat{x}_n(t) =
\sum_{\alpha =1}^\infty \biggl{[} \hat{b}_{\alpha} A_n^{\alpha} e^{-i
\omega_{\alpha} t } + \hat{b}^{\dag}_{\alpha} A^{ \alpha *}_{n} e^{i 
\omega_{\alpha} t  } \biggr{]},
\label{eq:displacement1}\eeqa
 with $\alpha$ being
the eigenvalue index, corresponding to the energy of the mode
$\omega_{\alpha} = \alpha \omega_0$ with amplitude $A_n^{\alpha}$
and $\hat{b}_{\alpha}$ being the mode annihilation operator. We
determine the breather solutions as described in~\cite{Cretegny}
and obtain the amplitudes $A_n^{\alpha}$ at site $ n $ along the
molecule. The coefficients $A_n^{\alpha} $
 describe the real space amplitude distribution  of the breather mode
with
frequency $\omega_{\alpha}$. This is
 a classical solution that will not decay in time (we ignore weak damping
to simplify the calculation).
Then 
\beqa &D^{0,r}(n,m;t,t^{\prime})  = -i\theta(t-t^{\prime})
\sum_{\alpha = 1}^{\infty} \nonumber \\ &
\times [A^{\alpha*}_nA^{\alpha}_m e^{ -i\omega_{\alpha}(t-t')}
 -A^{\alpha}_nA^{\alpha*}_m e^{ i\omega_{\alpha}(t - t')}]. & \;\nonumber \\
\label{EQ:displacement3} \eeqa 
Its Fourier transform is obtained
as: \beqa
 D^{0,r}(n,m;\omega) =  \sum_{\alpha = 1}^{\infty} [\frac{A^{\alpha *}_n
 A^{\alpha}_m}{\omega+\omega_{\alpha}+i\delta}
 -\frac{A^{\alpha}_n
 A^{\alpha*}_m}{\omega-\omega_{\alpha}+i\delta}]
  \label{EQ:dispalcemnt4}
 \eeqa
with infinitesimal $\delta$, from which the spectral function
follows immediately as
 \begin{equation}
 B(n,m;\omega) = \sum_{\alpha=1}^{\infty}
 A_{n}^{\alpha}A_{m}^{\alpha}[\delta(\omega+\omega_{\alpha})
 -\delta(\omega-\omega_{\alpha})]\;.
 \label{EQ:B1}
 \end{equation}
Here, without loss of generality, we have assumed $A_{n}^{\alpha}$ 
to be real. The electronic self-energy in 2D is then found to be: \beqa
&\Sigma(\br_n, \br_m;i\omega_{n}) =  \lambda^2 N_0 J_0(k_F r_{nm})
\sum_{\alpha=1}^{\infty} A^{\alpha}_n A^{\alpha}_m  &
\nonumber\\
&\times \int d\epsilon \biggl{[}\frac{1-n_{F}(\epsilon)
+n_{B}(\omega_{\alpha})}{i\omega_{n}-\epsilon-\omega_{\alpha}} -
\frac{1-n_{F}(\epsilon)
+n_{B}(-\omega_{\alpha})}{i\omega_{n}-\epsilon+\omega_{\alpha}}
\biggr{]}.& \label{eq:sigma2} \eeqa

\begin{figure}[th]
\centerline{\psfig{file=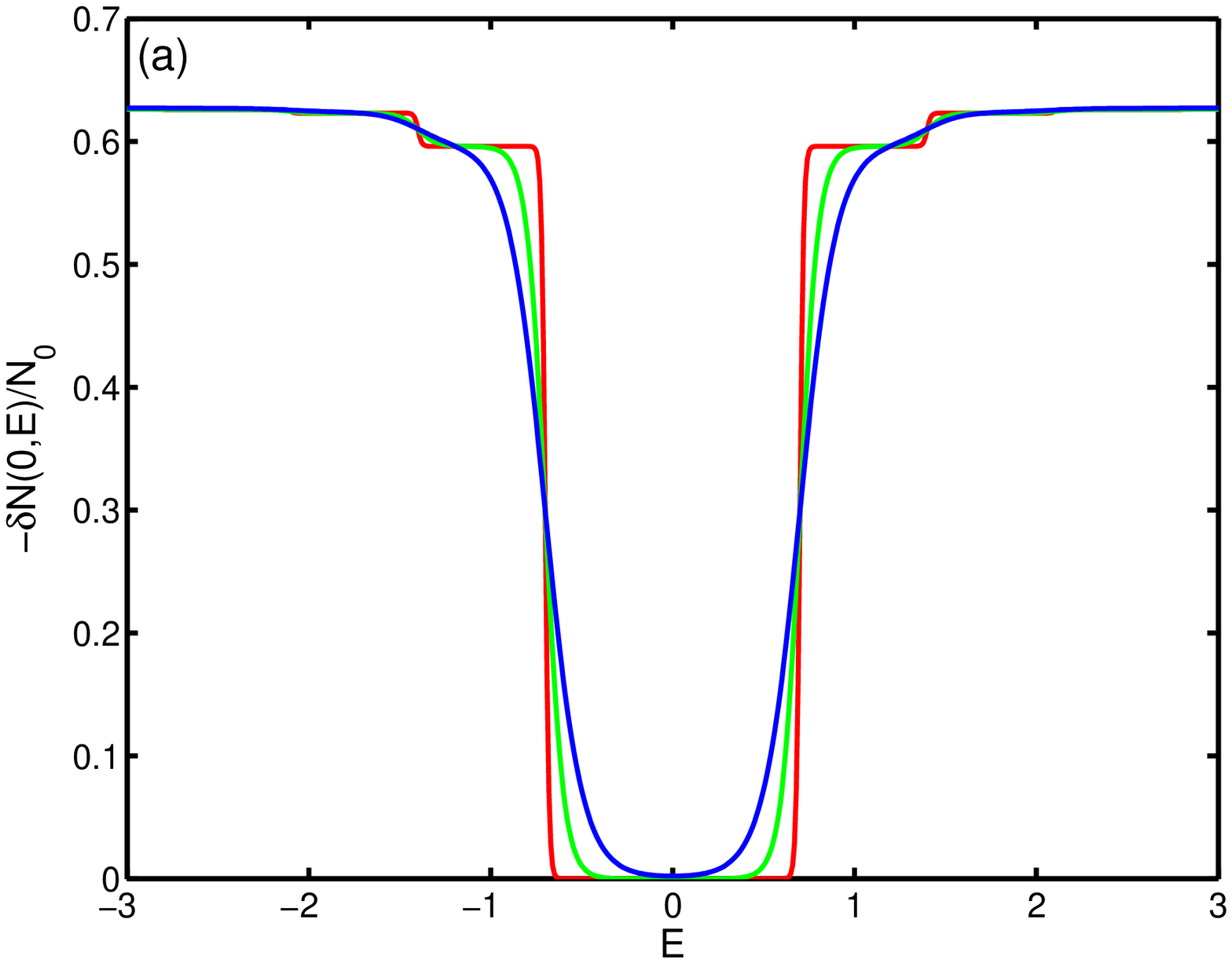,width=8cm}}
\centerline{\psfig{file=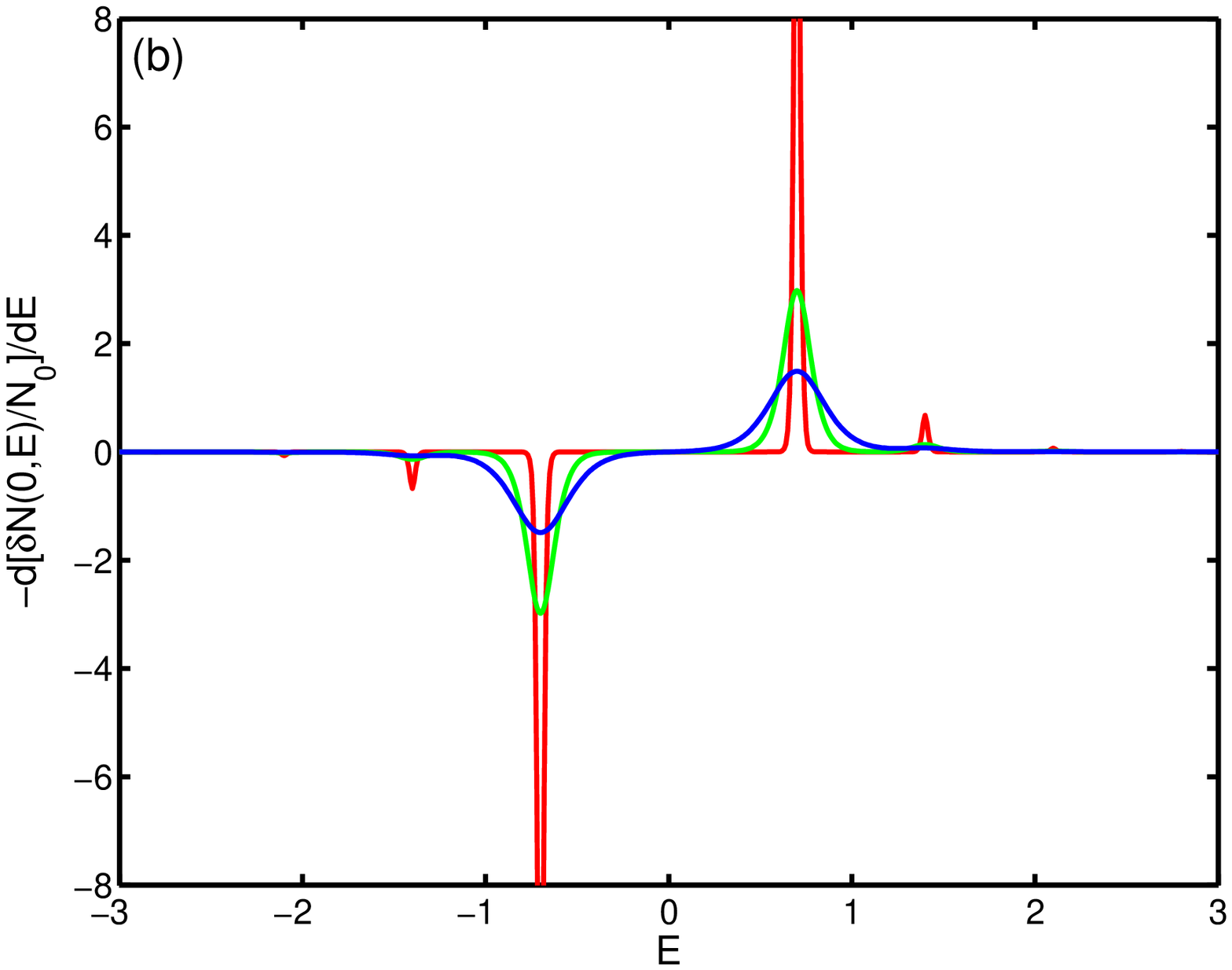,width=8cm}} \caption{(Color) The
correction of the local density of states calculated at the center
of the breather mode (a) and its derivative (b) as a function of
electron energy for various temperatures $T=0.01$ (Red), 0.05
(Green), and 0.1 (Blue).} \label{FIG:DOS}
\end{figure}
 
\begin{figure}[th]
\centerline{\psfig{file=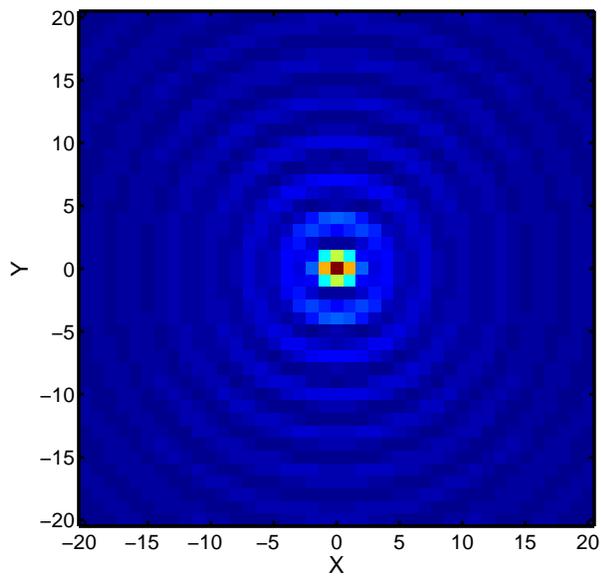,width=8cm}} \caption{(Color) The spatial
image of the derivative of the LDOS correction at energy $E=\omega_{0}=0.7$ 
due to the electron scattering off the breather mode. The temperatures
$T=0.01$. } \label{FIG:IMAGE}
\end{figure}

These formulas are quite general. The only breather-specific
features in Eqs.~(\ref{eq:sigma2}) and (\ref{EQ:B1}) are that the
breather amplitude factors $A^{\alpha}_n$ are localized at the
position, say $n = 0$, where the breather is excited~\cite{comment1}. 
 
Once we arrive at the 
the electronic self energy due to the
scattering of electrons off the breather mode that resides on the
molecule, we are ready to calculate the change in the LDOS of electrons, which 
is given by:
 \beqa
 N(\br_i,\omega) = -\frac{1}{\pi} \mbox{Im} G(\br_i, \br_i,E+i0^{+})\;,
 \label{eq:DOS1}
 \eeqa
 where the retarded Green's function is obtained through the analytic 
 continuation from Eq.~(\ref{EQ:FULL_GREEN})
  \beqa
 &G(\br_i, \br_i;E+i0^{+})  = G^0(0;E+i0^{+})
& \nonumber\\ &+ \sum_{n,m} G^0(\br_i - \br_n;E+i0^{+})
 \Sigma(\br_n,\br_m;E+i0^{+})
 & \nonumber \\ & \times G^0(\br_m -
\br_2;E+i0^{+})\;. &
 \label{eq:DOS2}
 \eeqa
The LDOS is proportional to the low temperature local tunneling conductance, 
which can be measured by STM~\cite{RJBehm1990}. 
 
A little algebra  leads to the correction to the LDOS: 
\beqa
&\frac{\delta N(\br_i, E)}{N_{0}} = -\pi^2 (\lambda N_0)^2
\sum_{n,m} \sum_{\alpha=1}^{\infty} A_{n}^{\alpha
}A_{m}^{\alpha}
& \nonumber \\
&\times J_0(k_F r_{1n})J_0(k_F r_{nm}) J_{0}(k_F r_{mi}) &
\nonumber \\
&\times [n_{F}(E+\omega_{\alpha})-n_{F}(E-\omega_{\alpha})
+ n_{B}(\omega_{\alpha}) -n_{B}(-\omega_{\alpha})] \;. \nonumber \\
\label{eq:DOS3} \eeqa
Equation ~(\ref{eq:DOS3}) is the central result of this paper.
It allows one to calculate the corrections to the LDOS at any point
on the surface and quantify the effects of the inelastic tunneling
corrections due to coupling to a breather. 

For the numerical calculation, we take the parameter values 
in Eq.~(\ref{EQ:Ham1}):
The atomic mass $M=1$, the elastic constant $k=0.1$, 
and a Morse potential $V(x)=
D[1-\exp(-x^2/\xi_{0}^{2})]$ with $D=0.5$ and $\xi_0=1$, so that the bottom 
of the phonon continuum is located at $\omega_{c}=1$. 
The lattice constant of the chain molecule is taken to be $\xi_0$.
Energy is measured
in units of $\omega_{c}$ and length in units of $\xi_0$.
We then choose the fundamental frequency of the 
breather mode $\omega_0=0.7$, and the dimensionless electron-mode 
coupling constant $(\lambda N_0)^2 = 10^{-2}$. As shown in 
Fig.~\ref{FIG:XDISTR}, the breather 
excitation is localized at some region in the molecule with a length scale of 
about four lattice constants along the chain direction, in contrast to the vibration 
mode of a single-atom molecule, which is located on a single site. Moreover, due to 
the anharmonicity of the Morse potential, the coefficients for higher frequency 
Fourier modes are finite. All these unique features can be detected by the 
IETS-STM experiments. Figure~\ref{FIG:DOS}
illustrates the correction to the LDOS (upper panel) 
due to the presence to the breather, and the 
effects on the derivative of the LDOS (lower panel). 
As expected, there are steps in the 
LDOS occurring at the energy $E=\omega_0$, $2\omega_{0}$, $3\omega_0$, \dots. 
For example the step at $E=2\omega_0$ originates from the inherent 
anharmonic nature of the 
breather but has much weaker intensity than the first step. However,  
corresponding to these steps, 
the peaks in the derivative of the LDOS are easier to identify.
These features are however rapidly smeared out as the temperature is increased.
Therefore, to detect the breather excitation, the temperature must be much 
lower than the mode frequency. Experimentally, the temperature of several 
Kelvins and a typical value of $\omega_{0} \sim 100 \mbox{meV}$ are 
easily accessible.  

Figure~\ref{FIG:IMAGE} shows the spatial distribution of the 
derivative of the LDOS correction at energy $E=\omega_0$. The strongest 
intensity occurs at the center of the localized breather core. 
However, one can also clearly see the 
spatial extension away from the breather 
(especially along the chain direction lying horizontally), which can be detected again 
by the IETS-STM. The ripples in the far field from the breather comes from
the rapidly oscillating factors $J_0(k_f r)$. All these features are observable 
by the IETS-STM by measuring the local tunneling conductance  $dI/dV$ and 
its derivative $d^2I/d^2V$. The strongest signature of the localized lattice
excitation is in the vicinity of the breather.

In summary, we have proposed a novel application of STM to perform IETS
on  extended molecules. Intrinsically localized nonlinear 
excitations (breathers) can  first be excited and
subsequently imaged using IETS STM. We find that indeed breathers
result in an extremely localized IETS features and may be
observed with STM at low temperatures. The formalism we have developed for
the IETS of an extended objects is quite general and is applicable
to other systems. One of the most interesting extended systems
where this experiment could be performed is IETS on DNA.
 
{\bf Acknowledgments:} This work was supported by DOE LDRD at Los
Alamos. We are grateful to Prof.  D. Campbell, Prof. T. Kawai and
Prof. H. Tanaka for useful discussions.


\begin{thebibliography}{99}
\bibitem{comment1} For the  case of
phonon excitations the problem becomes translationally invariant
and working in momentum space is more convenient.
\end{thebibliography}
\end{document}